# Identifying Terms and Conditions Important to Consumers using Crowdsourcing

XINGYU LIU, UCLA
ANNABEL SUN, JASON I. HONG, Carnegie Mellon University

Terms and conditions (T&Cs) are pervasive on the web and often contain important information for consumers, but are rarely read. Previous research has explored methods to surface alarming privacy policies using manual labelers, natural language processing, and deep learning techniques. However, this prior work used pre-determined categories for annotations, and did not investigate what consumers really deem as important from their perspective. In this paper, we instead combine crowdsourcing with an open definition of "what is *important*" in T&Cs. We present a workflow consisting of pairwise comparisons, agreement validation, and Bradley-Terry rank modeling, to effectively establish rankings of T&C statements from non-expert crowd workers on this open definition, and further analyzed consumers' preferences. We applied this workflow to 1,551 T&C statements from 27 e-commerce websites, contributed by 3,462 unique crowd workers doing 203,068 pairwise comparisons, and conducted thematic and readability analysis on the statements considered as important/unimportant. We found that consumers especially cared about policies related to after-sales and money, and tended to regard harder-to-understand statements as more important. We also present machine learning models to identify T&C clauses that consumers considered important, achieving at best a 92.7% balanced accuracy, 91.6% recall, and 89.2% precision. We foresee using our workflow and model to efficiently and reliably highlight important T&Cs on websites at a large scale, improving consumers' awareness.

## 1 INTRODUCTION

T&Cs often contain important information to consumers, such as an automatic subscription plan following a "free trial" [9], complicated cancellation or return policies [13, 25], and prevention of filing class action lawsuits [27]. However, past work has repeatedly shown that most consumers do not read or visit such policies [10, 13, 23, 42], because they do not believe that reading them will yield valuable knowledge or is a good use of their time [21]. Furthermore, eye-tracking data indicates those who do read policies usually stop after the first paragraph [52].

Prior work has explored methods to surface important information from T&Cs, but has multiple limitations. Terms of Service; Didn't Read (TOS;DR) [45] had legal experts review policies manually. However, it faces serious scalability issues, given the sheer amount of websites and frequent updates. It currently only covers about 450 websites with many of its analyses out of date due to changes to those policies. More recently, researchers have tried ML and NLP methods to summarize privacy policies. While privacy policies and T&Cs are different, the methods used could conceptually be adapted for T&Cs. Wilson et al. [20, 54, 55] had crowd workers used pre-determined labels (e.g. collects contact info, shares location info) to categorize privacy policies and trained ML models based on this crowdsourced data. However, this approach made two assumptions and would not work well for T&Cs: (1) The use of pre-determined labels for annotations imposed a strong assumption on what types of information the policies might contain. This might work for some privacy policies because they are regulated by laws (e.g. California's Online Privacy Protection Act) and are relatively more standardized (e.g. how companies collect, use, share, and delete each type of data). However, T&C documents contain a wider variety of content, which makes them harder to capture using pre-defined labels. (2) This past work is based on what researchers defined as important, and made many assumptions about what users care about, which may not be correct. T&Cs especially often include information closely related to *consumer behavior*, rather than legally binding contracts. Previous research has shown that in addition to legal risks and concerns, consumers also use T&Cs as a source of information, for copyright policies, specific return policies, and community rules [1, 14, 33].

To address these problems in the context of T&Cs, we chose to approach this problem from the consumers' perspective. We aimed to identify information that is relevant and important to the general consumers, and formatively understand what general consumers perceive as important in T&Cs. More specifically, in this paper, we seek to explore the following questions:

- How can we efficiently and reliably discover what kinds of information consumers consider important in T&Cs?
- What do consumers care about in T&Cs? Do consumers care about the same things?
- Can we build a scalable and accurate machine learning model to surface consumer-centered information in T&Cs?

To answer these questions, we first explored and examined the use of crowdsourcing, pairwise comparison, and Bradley-Terry rank modeling [5] to generate user-centered importance rankings for T&Cs individual sentences (we also use the term "statement" synonymously). We used this method to create a robust data set with 203,068 comparisons from 27 e-commerce websites' T&Cs, which is available for download[1]. We then validated that consumers have high agreement on what statements are generally deemed important. To examine what consumers care about, we conducted a thematic analysis of the top-ranked statements across all T&Cs, and found consumers to especially care about after-sale and monetary statements. We also observed that people tended to consider harder-to-understand statements as more important. Lastly, we built machine learning models to predict the within-policy importance of a statement from its text with balanced accuracy around 90%. Here, an ML model is useful because it offers new kinds of opportunities for surfacing important information in T&Cs to consumers, e.g. a website that aggregates policies across the web and lets people browse and see what our model highlights as "important", or browser plug-ins that can present useful just-in-time information (such as right before creating an account or right before purchasing an item).

In this paper, we make the following research contributions:

- A workflow to effectively and efficiently identify statements consumers considered *important* in T&Cs.
- Analysis and insights on what consumers find important in e-commerce T&C documents in our data set.
- ML models (SVM) that can identify important T&C clauses that are likely to be relevant and understandable to other consumers, with at best 92.7% balanced accuracy, 91.6% recall, and 89.2% precision.
- A data set of crowdsourced consumers' comparisons and rankings of what T&C statements are more important to them, consisting of of 1,551 statements from 27 T&Cs, and 203,068 pairwise comparisons.

## 2 RELATED WORK

In this section, we explain the differences between privacy policies and terms and conditions. We also outline past research on methods to make online policies easier to read, as well as voting methods to collect data from crowd workers.

---
[1]The crowdsourced data can be downloaded at *https://bit.ly/clearterms-data*

## 2.1 Privacy Policies vs. Terms and Conditions

Many websites have privacy policies (PPs), which tend to focus on providing specific information about how companies collect, use, share, and delete personal data of users. PPs are regulated by laws such as California's Online Privacy Protection Act. Terms and conditions (T&Cs) tend to contain multifaceted statements which largely depend on the kinds of services the website provides, and are more closely related to consumer behavior including billing and subscription, return policies, warranties, community rules, etc. Our focus in this paper is on T&Cs which are more diverse and consumer-centered, and less explored by researchers.

## 2.2 Alternatives to Terms & Conditions and Privacy Policies

Researchers have developed a variety of methods to assist people in understanding legal documents like T&Cs. We discuss some of the existing tools and their influence on our research.

P3P is a machine-readable language that lets websites share their privacy policies in a machine-readable format. However, P3P has seen little adoption, with only 10% of the most popular websites using it in 2008 [12]. Further, past research found that a large number of participating websites had incorrectly filled P3P policies [12, 26]. Another approach used privacy seals to indicate trustworthy websites, having websites submit their privacy policy to verification websites for approval, e.g., TRUSTe Trustmark, WebTrust, and BBB Online (no longer in operation). In theory, users could save time from reading privacy policies by seeing that the privacy seal exists, but previous work has indicated that these seals are unrecognizable to many people [3, 34]. This body of work also focuses on privacy, rather than Terms and Conditions.

Kelley et al. developed "privacy nutrition labels", which used P3P for displaying privacy policy information in a color-coded grid [24]. However, this method also only focuses on privacy, and risks missing important information that do not fit in the pre-defined categories. Balancing information reduction while retaining important information is a constant struggle with all summarization techniques.

Other approaches seek to restructure a policy's format to increase readability. Bartlett and Plaut presented T&C policies to users in bullet points with links to more detailed information [39], while Bernstein [4] and Waddell et. al [53] reorganized End-User License Agreements (EULA) by paraphrasing and presenting information across multiple windows. Their work shows the importance of incorporating visual design principles to increase readership, but is not actively informed by what the general consumer's top concerns and considerations for opting into a product or service are. In line with this concern, a recent experimental study by Ben-Shahar and Chilton found that user comprehension and behavior remained unaffected even after applying design principles on privacy policies, suggesting that policy restructuring is not a complete answer to helping users [2].

Similar to our work in using crowdsourcing and machine learning, Wilson et al. and Harkous et al. examined ML techniques to automate the annotation of privacy policies [19, 54]. However, this past work had crowd workers use pre-determined labels to annotate, which may not fully capture what consumers actually care about. The website Useguard[2] used a deep neural network to read privacy policies, but it only focused on identifying potential privacy threats, instead of general terms and conditions [50].

TOS;DR is a website dedicated to summarizing Terms of Service policies by manually examining policies. However, TOS;DR has high labor costs. Each volunteer must manually read the entire document, highlight each important sentence, and then explain the reasons for selecting it. While this method is viable for extracting important statements, there are issues with scalability, as the website has only analyzed about 450 policies since its inception in 2012. Our work extends their efforts by investigating an alternative crowd-based method tailored for speed and scalability.

---

[2]https://useguard.com

## 2.3 Voting Methods to Rank Items

There are several different voting methods for users to rank items. A common approach is to have voters select one or more candidates out of a pool of candidates, but such methods may be inappropriate for policy statements [37]. Policies can be lengthy, and having voters read all statements before making their selection incurs high labor costs and requires high memory retention. Other popular voting methods which do not require voters to read all statements employ a binary or an n-point Likert scale for individual items. This is a popular method among social science research as it allows data collection on a fine-grain and broad level. However, Chaianuchittrakul found that using binary and n-point Likert scales led to little or no differences in importance score distribution when crowd workers rated individual statements, making these methods inadequate for ranking policy statements [8].

Follow-up work by Angelia found that forcing users to select between two statements was more promising for ranking statements in privacy policies [1]. They used pairwise comparisons in a round-robin tournament setting, where winners of a comparison progress to the next round until a winner is found. This approach, however, yields comparisons that are dependent on previous matches, leading to: (1) an inability to analyze statement ranking independent of match history, and (2) a risk of statement mislabeling due to a poor matchup in early stages.

As such, we decided to use an all-pairs pairwise comparisons approach, where all possible pairings of items are compared against each other, resulting in a full list of all possible independent comparisons. Each pairwise comparison consists of selecting between two items at a time. Votes are aggregated for each item, and the item with the highest number of votes is the winner. The ability of pairwise comparisons to reduce the amount of text read and to sort highly subjective and unpredictable items from large data sets makes it a good candidate method for selecting important statements from T&Cs.

## 3 WORKFLOW DESIGN RATIONALE

Identifying what information consumers perceive as important in T&Cs is challenging because of the complexity and diversity of T&C statements and the current lack of understanding of what consumers care about in these policies. In developing a workflow to identify consumer-centered important information, we identified three major factors to consider:

**Efficiency**: Previous work shows that it takes on average 15–17 minutes for an adult to read through a typical T&C document [36]. We looked for ways of lowering the burden of work as well as how the work might be split up. Towards this end, we used crowdsourcing and divided the complex task of reading an entire T&C document into micro-tasks of pairwise comparison between two statements.

**Reliability**: Crowd workers should ideally generate reliable results. More specifically, we want crowd workers to make consistent decisions when reviewing and labeling T&Cs, especially given the open definition of *importance* and people's relatively low experience with such policies. We optimize this by structuring micro-tasks to be easy for workers to make decisions and having multiple workers do the same task. We also evaluated reliability by examining inter-rater agreement between workers.

**Scalability**: Due to the sheer amount of T&C documents and frequent updates of these policies, our proposed method needs to be scalable. That is, this method should be easily applied to other T&Cs with low cost. We improve scalability by applying a Bradley-Terry model to reduce the

Identifying Terms and Conditions Important to Consumers using Crowdsourcing

number of comparisons needed. We also built a predictive machine learning model to automatically identify T&C statements important to consumers for unseen e-commerce websites (see section 6).

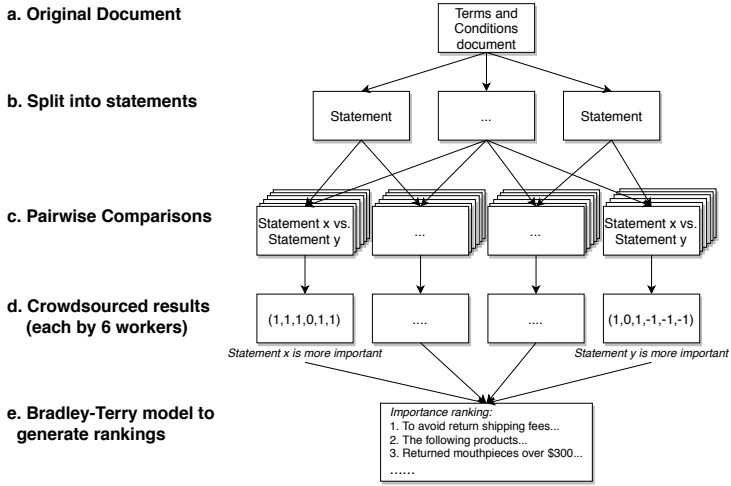

Fig. 1. Workflow of how we identify statements that consumers perceive as important in T&C documents. We first parse original T&Cs documents into statements (sentences) and combine them into pairs. Each pair is sent to 6 different crowd workers to judge which statement is more important (1 means that statement x is more important, 0 means equally important, and -1 means statement y is more important). We then aggregate the scores and fit a Bradley-Terry model using the results, to generate an importance ranking of T&C statements.

Below, we describe our workflow of using crowdsourcing, pairwise comparisons, agreement validation, and Bradley-Terry rank modeling, to identify important statements from T&C policies (Figure 1). We also offer more details about our design rationale and tradeoffs.

### 3.1 Open Definition of "Important"

We first need to determine what do we mean by *important*. One option is to ask users to match statements based on pre-defined criteria, similar to prior work on privacy policies. This approach would allow us to report reliable data for the criteria defined. Alternatively, we could let users choose which statements they find important, allowing us to report a wider variety of statements users find important and construct themes bottom-up. As mentioned earlier, because we do not know *a priori* what consumers find important in T&Cs, we opted to give our users an openly-defined task: *"Choose the statement that you think is more important for you to know, as a customer browsing the website."* (Figure 2) to learn what general consumers find important. This user-centered approach of leaving space for users to decide what is important also makes the task for crowd workers easier and more intuitive.

A concern here is whether crowd workers are consistent with each other in what they consider as important. This might affect the reliability of our method if workers have large disagreements. We found that even though we did not establish a formal definition for the word "important", different crowd workers still reached a high level of agreement in their judgements, as will be further discussed in Section 4.1.

## 3.2 Crowdsourcing Task Design

*3.2.1 Pairwise Comparison.* As noted earlier, previous methods [45, 54, 55] asked volunteers to read entire documents and flag relevant elements, which has serious challenges with efficiency and scalability. Other methods that asked crowd workers to label single statements are also not ideal here because they made strong assumptions on what content T&Cs contain and what consumers deem as important. We employed the idea of offering a pairwise comparison task where each worker compare two items at a time. Voting results from crowd workers can be aggregated together to produce the overall ranking, and thus finding the important T&Cs. This method not only makes the crowdsourcing tasks easier for each individual worker, but also allows participation from a larger population. In addition, research suggests that it reduces ambiguity and increases accuracy for people to compare between items than to draft a direct ranking or to evaluate each item alone (e.g. it can be difficult to give a direct ranking of or to score brands of wine, but may be feasible to compare a sample of pairs of wines and determine which one is better) [28, 56].

*3.2.2 Granularity of Information.* Pairwise comparisons in our case might not be wholly contained, that is, a piece of text in the T&Cs might refer to or depend on text mentioned elsewhere in the document. This naturally leads to the question of what is the right granularity for our comparison task. We considered having people compare paragraphs, but found that many T&Cs have rather long paragraphs, which would make the task inconsistent (e.g. comparing a short paragraph to a long one), difficult to compare, and time consuming. We also considered showing several sentences at a time, but had concerns that this would make the task harder and slower, for similar reasons as above. Furthermore, there was also no guarantee that any dependent text would be included. After considering the trade-offs, we opted to choose a granularity of one sentence to simplify the comparison task. We acknowledge that our approach will sometimes lead to a lack of context, but we wanted to start with the simplest interface. We found lengths of statements to be similar in the data we collected (Min = 1, Q1 = 14, Median = 22, Q3 = 32, Max = 59). Also, as we will discuss later in section 5.2.4, we found in our studies that crowd workers were consistent in judging important terms even without full context, e.g. *"bill forwarded to a collections agency if not paid in a month"* — vinotemp.com.

## 4 STUDY 1: IDENTIFYING T&CS IMPORTANT TO CONSUMERS

Using the task design mentioned in the last section, we conducted two data collections and studies as described in section 4 and 5. In this section, we describe in details how we collected data and established rankings of statements. In section 5, we provide insights on what consumers found important.

*4.0.1 Terms and Conditions Used.* For this initial study, we selected 12 T&C documents (with a total of 601 statements) of e-commerce websites servicing the US market (see Table 1). We chose e-commerce because we wanted a domain where sites would have somewhat similar characteristics, and because e-commerce websites have T&Cs more relevant to consumers (e.g., return policies, shipping policies, warranties) rather than generic legal contracts.

*4.0.2 Crowdsourcing Platform and Participants.* We used Amazon Mechanical Turk (MTurk) as our crowdsourcing platform to collect data from master workers who had a minimum of 95% acceptance rate for completed Human Intelligence Tasks (HITs) [40]. We also limited the location of MTurkers to be in United States only, since we are only evaluating e-commerce websites primarily serving the US market, and other cultures may interpret, understand, and find importance of T&Cs differently.

We paid $0.04 for every pairwise comparison workers completed, based on an estimate of minimum wage. Each HIT consisted of a randomly selected pairwise comparison from the pool of

Identifying Terms and Conditions Important to Consumers using Crowdsourcing

| Website | # Statements | # Votes per pair | # Comp. | Mean %Agreement | $[25^{th}, 75^{th}]$ %Agreement |
|---|---:|---:|---:|---:|---|
| Geekbuying | 44 | 6 | 5,676 | 0.73 | [0.67, 0.83] |
| GuitarCenter | 44 | 6 | 5,676 | 0.76 | [0.67, 0.83] |
| HomeDepot | 39 | 6 | 4,446 | 0.71 | [0.67, 0.83] |
| HP | 32 | 6 | 2,976 | 0.70 | [0.67, 0.83] |
| Lenovo | 78 | 6 | 18,018 | 0.73 | [0.67, 0.83] |
| Logitech | 33 | 6 | 3,168 | 0.74 | [0.67, 0.83] |
| Myntra | 42 | 6 | 5,166 | 0.71 | [0.67, 0.83] |
| Playstation | 77 | 6 | 17,566 | 0.75 | [0.67, 0.83] |
| Samsung | 70 | 6 | 14,490 | 0.69 | [0.5, 0.83] |
| Selfridges | 42 | 6 | 5,166 | 0.72 | [0.67, 0.83] |
| Uniqlo | 63 | 6 | 11,718 | 0.70 | [0.67, 0.83] |
| Zaful | 37 | 6 | 3,996 | 0.71 | [0.67, 0.83] |
| **Total** | 601 | 6 | 98,062 | 0.72 | [0.67, 0.83] |

Table 1. Detailed information of 12 T&Cs used for the initial crowdsourcing task.

all possible pairwise comparisons within the same T&Cs document. Each pairwise comparison was assigned to 6 unique crowd workers. Note that in our approach, we had 3 MTurkers comparing A vs. B (where A and B are statements from the policy), and 3 more comparing B vs. A to minimize ordering effects [48], which leads to a total of 6 pairwise comparisons for each statement. Crowd workers were allowed to complete multiple HITs.

For example, the T&Cs for homedepot.com contains 39 statements, resulting in 4,446 ($39 \times 38 \times 3$) pairwise comparisons, with a long tail distribution of contributions by crowd workers. Table 1 displays details about the companies. In total, this first study had 98,062 comparisons from 12 T&Cs from e-commerce companies.

Crowd workers were shown two randomly selected statements from the same policy as well as the source of the T&C policy, and asked to click on the statement they thought was more important to know (Figure 2). They could also select an option indicating that both statements were equally important. The resulting selection was converted into points: a statement received +1 points if selected, 0 points if selected as equal, and -1 points if the other statement is selected.

To prevent malicious random selections [17], we post-checked crowd workers' comparison results by manually examining votes for trivial comparisons (e.g. *"You will be fined $400 for returning damaged products."* vs. *"Our goal is your total satisfaction."*). No misbehavior was found.

### 4.1 Inter-rater Agreement Validation

Here, we examine to what degree different crowd workers agree on what they deem important. As noted earlier, higher consistency would suggest that consumers care about many of the same things, and offers support that our approach using pairwise comparisons has merit. In contrast, if there is low consistency, it would suggest that our method would have problems generalizing.

We evaluated the inter-rater reliability with percent agreement measurement [30, 47] of crowd workers' votes on each pairwise comparison (Table 1). Results showed that crowd workers had high agreement when performing pairwise comparison tasks (Figure 3), with a mean percent agreement score of 0.72 (more than 4 out of 6 different crowd workers voted the same). All comparisons were at least agreed by 3 out 6 crowd workers, 79.3% of the comparisons were agreed by at least 4 workers, 41.8% were agreed by at least 5 workers, and 13.4% received unanimous agreement. Thus, crowd workers have high agreement on what statements are important.

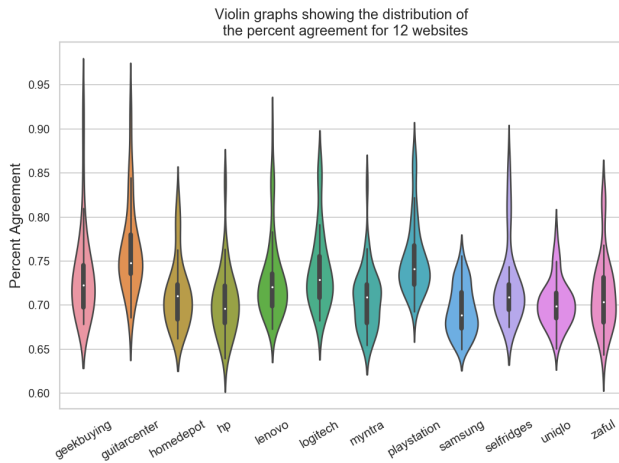

Fig. 2. Crowd workers were shown a pair of randomly selected statements from the same policy, and asked to select the more important of the two. Statements that were selected more frequently than others would be considered as higher in importance in our models.

Fig. 3. 12 websites' violin graphs of percent agreement between 6 crowd workers for each pairwise comparisons. Crowd workers reached high agreement for pairwise comparison tasks of selecting which statement is more important.

In addition to validating the inter-rater agreement, we also observed a trend that crowd workers had higher agreement when comparing statements that have larger discrepancies in importance. This correlation further validates both crowd workers' agreement on what is important and the soundness of our ranking model. This will be discussed in details in the next section.

Identifying Terms and Conditions Important to Consumers using Crowdsourcing

| Statement Text | Rank |
| --- | --- |
| To avoid return shipping fees, return your purchase at your nearest Guitar Center retail location, Limitations and Exceptions to Extended Return Period Returns of the following products must be made within 14 days of purchase... | 1 |
| The following products cannot be returned: Clearance items, items identified as non-returnable... | 2 |
| Returned mouthpieces over $300 incur an $8.00 sanitization fee; the fee for mouthpieces under $300 is $4.00 Apple products incur a 10% restocking fee on any open box returns.... | 3 |
| Our goal is your total satisfaction. | 42 |
| If you're not satisfied, neither are we. | 43 |
| Last Revised: February 22, 2018. | 44 |

Table 2. The top-3 and bottom-3 crowd-ranked statements for Guitar Center. Many of the top statements relate to finances and returns.

## 4.2 Building Rankings from Pairwise Comparisons

We then processed the pairwise comparison results by (1) aggregating the votes from crowd workers for each statements pair and (2) fitting a Bradley-Terry model to produce a ranking of statements. We chose to use the Bradley-Terry model because this probabilistic model can derive full rankings with sparser pairs [5], which can help reduce the number of crowdsourcing tasks and improve the scalability of our workflow. We further validated this and experimented how many tasks are needed in Section 4.3.

*4.2.1 Aggregating comparison results.* We aggregated the scores for the 6 individual HITs for each same pairwise comparison. Research shows that this majority vote approach to generate one annotation out of several opinions can filter noisy judgements and increase vote quality [35]. For example, let us assume that the results of the 6 hits for $statement_1 vs. statement_2$ are $[1, 1, 1, 1, -1, 0]$, which means crowd workers said $statement_1$ was more important than $statement_2$ four times, the opposite once, and equally important once. This would lead to an aggregated voting score of 3. If this sum is greater than 0, we label it as the first statement wins. If this sum is smaller than 0, then the first statement loses. If this sum is exactly 0, we remove this comparison from the data set. We then create a set of tuples for each of the T&C documents, based on these comparison results, representing the event "$statement_1$ wins over $statement_2$" as $(statement_1, statement_2)$. By convention, the first statement of the tuple represents the statement which wins.

*4.2.2 Fitting a Bradley-Terry model to generate rankings.* The Bradley-Terry model is a probabilistic model that can predict the outcome of a paired comparison [5]. It can be fitted to a particular set of pairwise comparison data [22], in our case the comparisons of T&C statements, and then derive a full ranking of the statements of that document. Specifically, this model assumes that in a comparison between two items $i$ and $j$, the probability that $i$ beats $j$ can be modeled with two positive parameters as $\frac{a_i}{a_j}$, which respectively represents the *ability* of $i$ and $j$. This probabilistic model especially has advantages in inferring the *ability* of items even when direct comparisons were not included, which is helpful to reduce the number of crowdsourcing tasks needed and increase scalability.

We used the *choix* library [29] to fit a Bradley-Terry model on our pairwise comparison data. Taking the aggregated comparisons as input data, *choix* computes the maximum-likelihood estimate of model parameters, and returns an array of ranked statements, where statements ranked at the top are consistently found to be more important from users' perspectives, and vice-versa for statements ranked at the bottom. Table 2 shows an example ranking of statements for one T&C policy.

*4.2.3 Vote vs. rank difference.* With the rankings generated, we tested two hypotheses as sanity checks of the quality of the crowdsourced comparisons and the ranking models: both the aggregated voting scores (e.g., $[1, 1, 1, 1, −1, 0] → 3$) and the percent agreement (e.g., $[1, 1, 1, 1, −1, 0] → 0.67$) of pairwise comparisons should have positive correlations with the rank difference between the two statements compared (note that rankings were generated only with tuples and individual votes were not used in fitting the Bradley-Terry models, thus this is not circular reasoning). That is, crowd workers should make clearer and more unanimous decisions when one statement is significantly important than the other. Two Pearson correlation tests on *aggregated voting score* vs. *rank difference* and *percent agreement* v.s. *absolute rank difference* both showed statistically significant correlations with p-values = 0 and < 0.001. The *aggregated voting scores* showed a stronger correlation with Pearson coefficient = 0.68, and the *percent agreement* has a coefficient = 0.29.

## 4.3 Scalability

While using an all-pairs pairwise comparisons approach gives us rich data, this method can be costly in terms of time and money since the number of pairwise comparisons grows $O(N^2)$, where $N$ is the number of statements in a policy. Reducing the amount of data needed for each policy can make it faster and cheaper to collect data, but there may be trade-offs when compared to the full data set, with respect to consistency of results and accuracy.

The Bradley-Terry model we used should be less vulnerable to these problems. In this section, we wanted to validate that and examine the tradeoffs between collecting less data and accuracy (as compared to the full data set). We tested the model's scalability by sampling subsets of increasing sizes from our full data set. For each data subset size, we compared the similarity between the subset-generated ranking and the ranking generated from the full data set, using Kendall's tau $\tau$ statistic [31] and percentage of overlap of the top-ranked statements.

We performed 10 rounds of data sampling from the full data (12 websites T&Cs statements from the previous study), beginning with sampling 10% of the pairwise comparisons, and incrementing that amount of data by 10% until we reached the full 100%. We began each round by randomly sampling comparisons, while ensuring all statements were represented in the subset at least once to avoid statements without any comparisons. Each subset was then fitted using the Bradley-Terry model as discussed previously to generate rankings of statements. We then calculated the Kendall tau $\tau$ correlation between the two rankings, as well as conducting a $\tau$ test. A small p-value in the test would show evidence of a monotonic relationship: a lower rank in the ranking generated from the subset is associated with a lower rank in the ranking generated using full data. The $\tau$ coefficient indicates the strength of such an association. Conventionally, $|\tau| > 0.35$ indicates a strong association and $|\tau| > 0.2$ indicates a medium association [51]. Table 3 shows that the $\tau$ coefficients for most websites become statistically significant and indicate at least medium associations by using around 50-60% of the data. This means that the rankings generated using half of the comparisons and more would be similar to the rankings generated using the full set. Since subsets were sampled randomly, we observed some difficulties in converging (e.g. Uniqlo) and gaps (e.g. Playstation at 60 and 70 percent). Thus, we further analyze how many of the top statements were captured with less data.

Since our goal is to identify the most important T&Cs statements, we mostly care about the model's ability to correctly capture the top-ranked statements. We performed an additional analysis to measure the percentage of statements in common between the top 20% ranked statements over a sample and the top 20% ranked statements over the entire data set (we call this *similarity coefficient*

Identifying Terms and Conditions Important to Consumers using Crowdsourcing

|  | %Comparisons used (Coef.) | | | | | | | | |
|---|---|---|---|---|---|---|---|---|---|
| **Website** | 10% | 20% | 30% | 40% | 50% | 60% | 70% | 80% | 90% |
| Geekbuying | 0.03 | 0.25* | 0.02 | 0.14 | 0.19* | **0.27**** | **0.30**** | **0.37***** | **0.43***** |
| GuitarCenter | 0.16 | 0.13 | -0.13 | 0.16 | **0.37***** | 0.25* | 0.21* | **0.48***** | **0.67***** |
| HomeDepot | 0.14 | 0.14 | -0.01 | -0.02 | **0.28*** | 0.20* | **0.30**** | **0.50***** | **0.65***** |
| HP | 0.09 | -0.05 | -0.06 | 0.05 | 0.016 | **0.22*** | **0.22*** | **0.53***** | **0.41***** |
| Lenovo | 0.12 | 0.03 | 0.21 | 0.15* | 0.16* | 0.14* | **0.20*** | **0.20**** | **0.35***** |
| Logitech | 0.11 | 0.10 | 0.08 | 0.41 | **0.29*** | **0.24*** | 0.17* | **0.34**** | **0.39**** |
| Myntra | -0.05 | -0.07 | 0.15 | **0.32**** | **0.23*** | 0.14* | **0.20*** | **0.29**** | **0.25*** |
| Playstation | 0.07 | 0.06 | 0.06 | 0.03 | **0.23**** | 0.15* | 0.16* | **0.43***** | **0.36***** |
| Samsung | 0.01 | -0.04 | -0.02 | 0.11 | 0.11 | **0.23**** | **0.22**** | **0.23**** | **0.24**** |
| Selfridges | -0.07 | -0.08 | 0.11 | **0.27*** | 0.17 | **0.20*** | **0.27*** | **0.21*** | **0.35**** |
| Uniqlo | -0.14 | 0.05 | -0.01 | 0.08 | 0.09 | 0.09 | 0.17* | 0.17* | 0.19* |
| Zaful | 0.17 | 0.04 | 0.19 | 0.14 | **0.24*** | **0.33**** | **0.25**** | **0.37**** | **0.63***** |

Note: *$p < 0.05$; **$p < 0.01$; ***$p < 0.001$

Table 3. Table showing the Kendall's $\tau$ coefficients and p-values for the rankings generated from 10 subsets vs. the complete set, for 12 websites. $\tau$ values above 0.2 (medium association) and are statistically significant are bolded and highlighted. Data started to show medium associations by using around 50-60%of the full data.

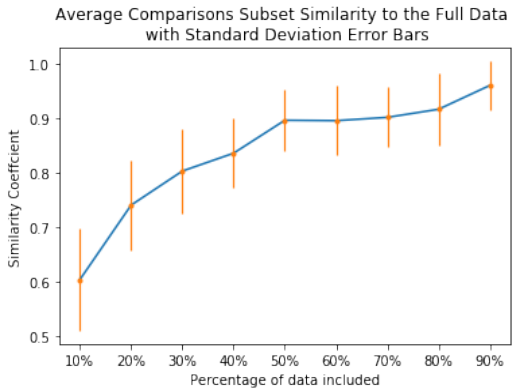

Fig. 4. This graph shows how well the rankings generated from subsets of the crowd can capture the top statements compared to the full crowd, averaged over all websites, with standard deviation error bars. Only including half of the data would yield a 90% similarity of the top statements.

synonymously). Our selection of 20% is a judgment call, and we consider it reasonable given we wanted to balance coverage with usefulness. We conducted 100 simulations for each of our T&C policies, where each simulation consisted of 10 rounds (going from 10% to 100% of the full data set). Afterwards, we averaged the similarity coefficients of all 12 T&Cs and summarized them in a line graph (Figure 4). It shows the overlap quickly converges towards the same top-ranked statements even with just 20-30% of the total data, and reaches at least 0.8 with mean value around 0.9 at 50% of the total data.

| Website | #Statements | #Votes per pair | #Possible Comp. | #Comp. |
|---|---|---|---|---|
| Neiman Marcus | 32 | 6 | 2,976 | 1,488 |
| CDW | 33 | 6 | 3,168 | 1,584 |
| PC Connection. | 35 | 6 | 3,570 | 1,785 |
| Macy's | 46 | 6 | 6,210 | 3,105 |
| American Girl | 52 | 6 | 7,956 | 3,978 |
| Vitacost | 55 | 6 | 8,910 | 4,455 |
| Sony | 56 | 6 | 9,240 | 4,620 |
| GameStop | 57 | 6 | 9,576 | 4,788 |
| VinoTemp | 61 | 6 | 10,980 | 5,490 |
| Toys'R'Us | 65 | 6 | 12,480 | 6,240 |
| Grainger | 69 | 6 | 14,076 | 7,038 |
| Nordstrom | 69 | 6 | 14,076 | 7,038 |
| CafePress | 73 | 6 | 15,768 | 7,884 |
| Amazon | 117 | 6 | 40,176 | 20,358 |
| NG Cleansing | 130 | 6 | 50,130 | 25,155 |
| **Total** | **950** | 6 | 210,012 | **105,006** |

Table 4. Breakdown of 15 more T&Cs used to build the machine learning model. #Possible Comp. is the number of all possible pairwise comparisons, and #Comp. is the number of comparisons we did for data collection. We reduced the amount of work by 50% by randomly sub-sampling from the full data set, and fitted Bradley-Terry models based on that.

## 5 STUDY 2: WHAT CONSUMERS CARE ABOUT IN TERMS AND CONDITIONS

Here, we present our analysis of statements that the crowd found important/unimportant. Knowing the content of these selections can provide a sanity check of our prediction models, and offer insight as to what consumers care the most about.

### 5.1 Collection of More T&Cs

We used the techniques discussed in section 4.3 and collected 15 more e-commerce T&Cs, yielding an additional 105,006 comparisons. Applying our findings, we optimized the data collection process by 50% (Table 4). Our final combined data set consists of 1,551 statements from 27 T&Cs, ranked from 203,068 pairwise comparisons. This complete data set is publicly available for download (see Footnote 1). We use this combined data set for the remaining studies.

### 5.2 Thematic Analysis of Important T&Cs

We collected the top 10% ranked statements for each T&C document from 27 websites (a total of 155 statements). We conducted a thematic analysis using Braun and Clarke's reflexive approach [6], to first create open-codings and lower level themes, and then inductively develop higher level themes and insights from codes (a statement can have multiple themes or codes). Our final code book consisted of 7 high-level themes (Table 5), and 24 lower-level codes (e.g. warranty eligibility, non-refundable, price change, etc.). In this section, we summarize findings from our thematic analysis.

*5.2.1 After-sales.* Statements related to after-sales, specifically return, repair, replacement and warranty appeared the highest number of times, comprising nearly half of the top 10% statements. We found consumers especially care about such service's eligibility, exceptions or extra costs. People wanted to be informed about special cases in these after-sales clauses. For example, guitarcenter.com includes a return policy that charges extra fees for some special products:

"*Returned mouthpieces over $300 incur an $8.00 sanitization fee; the fee for mouthpieces under $300 is $4.00 Apple products incur a 10% restocking fee on any open box returns.*" — guitarcenter.com

Geekbuying.com stipulates an exception for their free repair rules:

"*GeekBuying will be in charge of all the repair handling fees and non-artificial defective component replacement charges (this free repair rule does not apply to motherboard & screen)*" — geekbuying.com

Similarly, logitech.com explains about a special case of their return policies during the holiday season:

"*Each year we extended our normal 30-day return policy so that the holiday gifts you purchase from November 15 through December 23 can be returned for any reason until January 31 of the following year.*" — logitech.com

*5.2.2 Monetary Statements.* 47 of the top 10% important statements were about monetary including fee charge, credit, price change, and refund policies. For example, NG Cleansing contains a statement that charge consumers automatically after a free trial:

"*Unless you cancel within 14 days from today, you will be automatically charged the full purchase fee ($88.92) 14 days from today and enrolled in our auto-ship program.*"

Another commonly appeared type of statements we did not expect to see was statements related to price changes or price errors and websites' policies on how consumers would be compensated. For instance:

"*Should we lower the price of the product you purchased, you may contact us by email at myhp-sales@hp.com within the products return window to request a credit for the difference between the price you paid and the current HP.com Store selling price.*" — hp.com

*5.2.3 Other Important Statements.* 25 of the top 10% important statements were related to order changes, for example, logitech.com describes the conditions that the orders cannot be canceled: "*Orders for in-stock items are sent to our warehouse for shipping immediately after you place the order, therefore, in-stock items cannot be cancelled.*" Consumers also found statements related to Liability of damage and injury (e.g. "*LENOVO SHALL NOT BE LIABLE FOR MORE THAN THE AMOUNT OF ACTUAL DIRECT DAMAGES SUFFERED BY CUSTOMER, UP TO THE AMOUNT CUSTOMER PAID FOR THE PRODUCT OR SERVICE.*" — lenovo.com), customer service and communication (e.g. "*Email us with pictures attached of the received items/package and detailed information to describe the issue...*" — geekbuying.com), and terms amendment to be important.

*5.2.4 Statements without Context.* As mentioned in section 3.2.2, we also observed 9 statements that did not have contexts that were ranked in the top 10% important statements. For example,

"*If items of the above categories are faulty for non-artificial reasons within one year warranty (counting from the date of arrival), customers can send the items back for free repair.*" — geekbuying.com

*5.2.5 Unimportant Statements.* Among the bottom statements, we found content that do not appear to give consumers important or relevant information. We found business clichés, e.g., "*NOW GO ENJOY OUR SITES!*" (Playstation bottom rank #1), external links to other information, e.g., "*For

| Theme | #Statements |
| --- | --- |
| After-sales (Return, Repair, Replacement, Warranty) | 74 |
| Monetary Statements (Fee, Credit, Refund) | 47 |
| Order Changes | 16 |
| User Content and Behavior | 15 |
| Liability of Damage and Injury | 14 |
| Customer Service and Communication | 7 |
| Terms Amendment | 3 |

Table 5. High-level themes in the top 10% statements for all T&C documents from 27 e-commerce websites

*information regarding your intellectual property rights..."* (CafePress bottom rank #1), and others, e.g., *"This website is accessible to you at no cost."* (GameStop bottom rank #1).

## 5.3 Other Characteristics of Important T&Cs

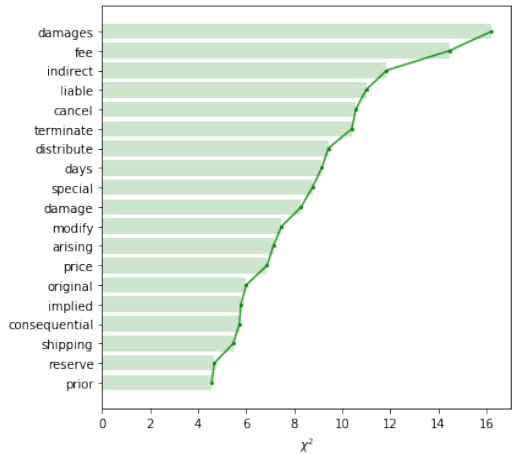

Fig. 5. Chi-square scores of words that correlate with important statements

*5.3.1 Words' Correlations with Importance.* We used Pearson's chi-squared test [11] to find the top 20 words most correlated with important statements in our data set (Figure 5). This plot suggests consumers place high value on potential financial loss, with terms like "fee", "price", and "damages". Many words are also related to return/cancellation policies, e,g, "cancel", "terminate", and "original", as are words relating to shipping polices, e.g. "distribute", "days", and "shipping".

*5.3.2 Readability vs. Importance.* Many prior work examined the readability of T&Cs and found them incomprehensible [16, 44]. A recent survey of 150 popular websites found that the vast majority of these privacy policies exceed the college reading level. We conduct this additional analysis to see if there is correlation between a statement's readability and its crowd-ranked importance, that is, do consumers found more difficult-to-understand statements more important, or vice versa?

Identifying Terms and Conditions Important to Consumers using Crowdsourcing

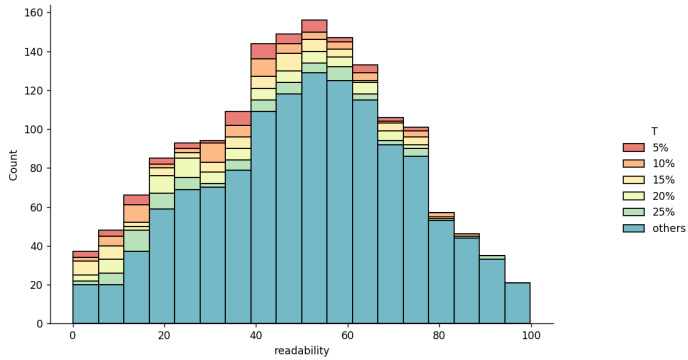

Fig. 6. A stacked bar plot showing the distribution of T&Cs statements' Flesch reading ease readability scores. Each bar is color-coded and stacked based on $T$, that is, the top-$T$% ranked statements.

We computed readability for statements using the Flesch Reading Ease score [15], which is on a scale from 0 - professional (extremely difficult to read), to 100 - 5th grade (very easy to read). Figure 6 shows a stacked bar plot of statements' readability scores and their ranks. We observed that most statements that had a readability score greater than 80 - 7th grade (fairly easy to read) were considered unimportant by consumers, while nearly half of the statements with a readability score < 20 - college graduate (very difficult to read) were in the top 25% of importance rankings. To further validate this correlation, we conducted a Kendall's $\tau$ test of statements' readability scores and their relative rankings (rank divided by total number of statements in that T&Cs). We obtained a $\tau$ coefficient = 0.22, with p-value < 0.001. This provides statistically significant evidence that there is at least medium association between the readability and the importance perceived by consumers of a statement. Consumers tend to regard statements that are harder to understand as more important.

## 6 PREDICTIVE MODEL

One extended application of our work is to use the crowdsourced raw data and rankings to predict the importance of T&Cs statements in other policies. As mentioned in the intro, one application would be to offer a web site that aggregates policies across the web and see what statements our model deems "important".

We trained machine learning classification models on the collected ranking data, where statements were divided into classes of "important" and "unimportant", and use the model to automatically identify which statements might be important to consumers, again only in the context of e-commerce web sites.

### 6.1 Model Building and Training

We first binned our data into 2 classes, "important" and "unimportant" by their importance rankings generated from crowdsourcing results. We chose classification of important/unimportant statements over regression because rankings/scores were generated comparatively within the context of each document. Thus, it was not possible to analyze the overall distribution and determine a threshold of important/unimportant with statistical evidence (e.g., a statement with a rank of 5 in Uniqlo's T&Cs is not necessary more important than a statement with a rank of 7 in Geekbuying's T&Cs). Specifically, we label the top $T$% of ranked statements of each website as "important", and the rest as "unimportant". The selection of the importance threshold, $T$, is debatable and we considered it as a judgement call that people can modify to adjust the sensitivity and precision/recall of their models, similar to participatory ML systems like [18]. In this study, we experimented with different threshold settings for $T$, and reported their performances.

*6.1.1 Data Pre-processing.* Before we start the training process, we applied some conventional text pre-processing steps to our data, including: converting to lowercase, removing punctuation, removing stop words, and lemmatization. In earlier iterations, we found that all caps and punctuation (such as exclamation points and dollar signs) did not suggest importance, contrary to our expectations. Hence, we removed those to make the ML models simpler. As mentioned in section 3.2.2, we also found lengths of statements to be similar, thus did not include that in the model.

*6.1.2 Text Representation: Word2Vec and tf-idf.* To produce more informative text representations, we used word embeddings, specifically Word2Vec, to capture and model semantic relationships between vocabularies. Since this model is unsupervised, high-quality word embeddings can be trained from a much larger set of unlabeled T&C statements [32]. In our study, we trained a customized Word2Vec model with 17,321 vocabularies from 427,814 T&Cs statements collected from the internet, and applied it onto our 1,551 labeled statements.

Specifically, we mapped each word to a 100-dimensional vector, and built features for each statement by multiplying word vectors with weights. We used tf-idf weighting scheme, which assigns each term ($t$, in our case each word in the statement) in a document ($d$, in our case the T&C statement) a weight based on its term frequency ($tf$, how many times the word appears in the sentence) and inverse document frequency ($idf$, the ratio of number of documents to the number of sentences in the corpus containing the term $t$) [43].

*6.1.3 Classification algorithm: SVM.* We used the pre-processed data to train a machine learning classification model. We selected the Support Vector Machine (SVM) algorithm and found it to perform the best in general after trying multiple ML algorithms including Logistic Regression, Naive Bayes, and Random Forest. We used Scikit-Learn's svm.SVC API [38] to construct and train the SVM classifier on our processed statements data, each with a label "important" or "unimportant".

*6.1.4 Training and Testing.* To avoid data contamination in the binning process, we fixed the unimportant statements as the statements that were ranked in the lower 50% in each website, and adjusted the importance threshold in the upper half. We then split the data into training and validation sets with a ratio of 8:2. Then, on the training set, we optimized the hyper-parameters with grid search, using Scikit Learn's GridSearchCV API [38], on $C$ (penalty parameter of the error term), $kernel$ (which kernel to use) and gamma $\gamma$ (the kernel co-efficient). Since this is an imbalanced data set, we used the balanced accuracy [7] as the metric when doing grid search, to avoid model bias. In grid search, models were tested with a 5-fold cross validation method. After finding an optimized set of hyper-parameters, we recorded the performances of the final model on the left-out validation set. We bootstrapped the above process 10 times for each threshold and reported the averaged metrics.

## 6.2 Results

*6.2.1 Model Performance.* We experimented with 5 different thresholds for binning the data, from the top-5% to top-25%, and applied the above pipeline on each data set with different threshold for labeling. For each model, we reported its balanced accuracy, recall, precision, and hyper-parameters used (Table 6). Most models showed good performances, while the thresholds of 10% and 15% performed the best with balanced accuracy of 93.4% and 92.7%.

Identifying Terms and Conditions Important to Consumers using Crowdsourcing

| $T$ | Balanced Acc. | Recall | Precision | Hyper-parameters |
|---|---:|---:|---:|---:|
| 5% | 87.8% | 73.9% | 94.3% | $C = 10, \gamma = 0.001, kernel = RBF$ |
| 10% | 93.4% | 90.0% | 90.0% | $C = 10, \gamma = 0.001, kernel = RBF$ |
| 15% | 92.7% | 91.6% | 89.2% | $C = 100, \gamma = 0.001, kernel = RBF$ |
| 20% | 90.2% | 91.8% | 87.2% | $C = 100, \gamma = 0.0001, kernel = RBF$ |
| 25% | 89.1% | 87.1% | 90.4% | $C = 10, \gamma = 0.001, kernel = RBF$ |

Table 6. Machine learning (SVM) models built with different binning thresholds ($T$) for "importance" and their performances. $T$ is the importance threshold, meaning that for this test the top $T$% of the ranked statements are labeled as important.

In addition, this model can be used to highlight important statements in T&Cs documents, simply by applying the predictive model to each of the statements and report the classification results. We tested our model (importance threshold = 15%) on the T&Cs statements of apple.com, cabelas.com and rei.com, as shown in Table 7. We can see that statements closely related to consumers' interests, for example, refund, order and pick-up policies, are classified as important to consumers.

*6.2.2 Error Analysis.* Our models perform well, but not perfectly. To understand its limitations and cases where it went wrong, we conducted an error analysis on the False Positive and False Negative predictions of the model.

**False Positives:** We observed that lots of the false positive predictions were statements that contained some sensitive keywords (e.g. "purchase", "price", "right", "charge" etc.), but did not necessarily convey important information, for example, *"You may not export any products purchased at HP.com."* — hp.com, and *"It is recommended, but not required to also purchase insurance or delivery confirmation services."* — amazon.com. This is probably due to the nature of our algorithm which largely depends on word embeddings. We also noticed a special case for this error: the model is predicting lots of "refund goes to original payment method" kind of statements as important, while crowd workers considered them unimportant. This is probably also caused by keywords like "refund", while most of them are common-sense to consumers. Examples include:

- *"Alipay orders returned to store will be refunded to your Alipay account."* — lenovo.com
- *"eVoucher orders returned by post will be refunded to a Selfridges eVoucher."* — selfridges.com
- *"Refunds will be issued based on the original form of payment."* — uniqlo.com

There are also false positive predictions which we did not see any pattern: *"In the event of substantive changes to this TOS, you may be notified by email."* — cafepress.com; *"By using Amazon Services, you agree to these conditions."* — amazon.com.

**False Negatives:** In general, false negative predictions have less in common. We only observed one pattern where statements were domain-specific and were not represented in the training data set. For example, NG Cleansing, an e-commerce website that mainly sells cosmetics contains policies related to health and medical conditions: *"If you are pregnant, nursing, taking prescription medication, have a history of heart conditions or have any medical condition, we suggest consulting with your doctor or primary care physician before using this Product."*

## 7 LIMITATIONS AND FUTURE WORK

**T&Cs consumers "care about" vs. "should care about":** We acknowledge that while both are important, what consumers "care about" and "should care about" are not necessarily the same and serve different purposes. In early explorations, we asked four legal experts, including two experts in corporate law and two experts in privacy working at a non-profit organization, to analyze two T&Cs that crowd workers had already ranked (Cabela's and Victoria's Secret). We found a high level of disagreement both between the experts themselves and between experts and crowd workers, similar to what was previously discovered in privacy policies [41, 54]. However, as we discussed in section 4.1, we saw agreement among crowd workers. While it is hard to define what is "legally" important in T&Cs, our consumer-centered approach can at least provide consumers suggestions of what other consumers considered important.

| Website | Predicted Important Statements |
|---|---|
| apple.com | In the event we cannot supply a product you ordered, Apple will cancel the order and refund your purchase price in full. |
| | Furthermore, there may be occasions when Apple confirms your order but subsequently learns that it cannot supply the ordered product. |
| | Apple is not responsible for actions taken by the third party once your item(s) have been picked up. |
| | Apple may, in its sole discretion, refuse or cancel any order and limit order quantity. |
| | In addition, your bank or credit card issuer may also charge you foreign conversion charges and fees, which may also increase the overall cost of your purchase. |
| cabelas.com | IN NO EVENT SHALL CABELA'S BE LIABLE FOR ANY SPECIAL, INDIRECT, PUNITIVE OR CONSEQUENTIAL DAMAGES RESULTING FROM ANY USE OR PERFORMANCE OF OR CONTENT ERRORS OR OMISSIONS IN THE INFORMATION, EVEN IF NOTIFIED IN ADVANCE OF THE POTENTIAL FOR SUCH DAMAGES. |
| | Any reproduction of these pages for commercial purposes or for distribution to other persons is a violation of United States Copyright law and may subject you to civil and criminal penalties. |
| | Trademark NoticeAll product names, trademarks, service marks or other images in this web site are either the property of, or used with permission by, Cabela's, and the use thereof without the express written consent of the owner(s) thereof is strictly prohibited. |
| | (1) Your submissions and their contents will automatically become the property of Cabela's, without any compensation to you; |
| | (2) Cabela's may use or redistribute the submissions and their content for any purpose and in any way; |
| | (4) there is no obligation to keep any submission confidential. |
| rei.com | REI Store Garage Sales are events where we sell merchandise as is. All sales are final on items purchased at REI Garage Sales at our retail locations. |
| | REIs guarantee doesnt cover ordinary wear and tear or damage caused by improper use or accidents. |
| | If your item has a manufacturing defect in its materials or workmanship, you can return it at any time. |

Table 7. Predicted important statements of 3 selected T&Cs (apple.com, cabelas.com, and dji.com) using our machine learning model.

**Granularity of comparisons:** As discussed earlier, one tradeoff with our approach is with our choice of granularity, namely individual sentences. We felt that individual sentences were a reasonable starting point, but it may be the case that some sentences refer to others, or cannot be easily understood without more context. There are several opportunities here for future work,

Identifying Terms and Conditions Important to Consumers using Crowdsourcingincluding trying other granularities, and using NLP techniques to automatically capture all related texts of each statement (e.g. text similarity or co-reference resolution).

**Long tail distribution of contributors:** We also had a long tail distribution of contributions from MTurkers. While this distribution does match that of other collaborative environments (e.g., All Our Ideas, Wikipedia, and studies on MTurkers has found the existence of "power users" who contribute a large part of work), we acknowledge the potential for bias [46, 49].

**Focus on e-commerce websites:** While we collected over 200,000 comparisons, it was only for 27 T&C policies and only for e-commerce websites. Though this amount of data was sufficient to perform experiments and analyze the method itself, another opportunity for future work is to gather more policies from and build models for domains outside of e-commerce (e.g. social media).

We also investigated one way of reducing data collection costs, and there are many other optimizations that remain to be explored. Three examples include reducing the number of pairwise comparisons dynamically based on worker agreement, creating ranking algorithms to determine which pairwise comparison might yield the most information gain, and creating a classifier that can find statements very likely to be ranked near the very bottom for importance. We also observed that there are some statements that are copies of or near-exact copies of each other, likely due to T&C policies being copied and pasted across different web sites. This kind of duplication might also help with scalability.

**Better ML models:** Our current ML models are based on word embeddings. While these show promising performance, we suspect that it also causes some false positive cases based on certain keywords, as discussed in section 6.2.2. To address this problem and further improve our model, we plan to experiment with more complex language models that can capture sentence-level meanings such as BERT.

**Applications of this work:** We are currently in the process of building out a website to showcase the results of our analysis on a large set of Terms and Conditions from e-commerce web sites, offering new kinds of interfaces and visualizations to surface important information from these policies to users. Also, as noted earlier, an alternative user interface might be to offer a browser plug-in that can highlight important T&C statements as users go through websites, offering relevant context-specific information just in time. We are planning to conduct further user tests to validate consumers' perception of important statements in a situated context, and evaluate how these approaches can influence consumers' consumption of T&Cs.

## 8 CONCLUSION

In this paper, we evaluated using crowdsourcing for pairwise comparisons of statements in Terms and Conditions policies, with an open definition of "importance". We have also made this data available for download (see Footnote 1). We used the Bradley-Terry model to aggregate these comparisons and rank statements. We also found that crowd workers had high agreement on what statements they considered important, showing that even with an open definition of "importance", crowd workers still generate consistent and reliable results. We then examined the method's reliability by testing data subsets of different sizes, and found that the top ranked statements converged and remained stable when including only 50% of all possible pairwise comparisons, suggesting that this method has potential to be an efficient approach. We presented an analysis of crowd preferences, which offered some insights towards users' concerns on T&Cs. In our thematic analysis, we found consumers to especially care about statements about special cases of after-sale policies and monetary statements. Lastly, we built working prediction models that can serve as a basis for tools and interfaces that can help consumers surface T&Cs they might consider important.

## 9 ACKNOWLEDGEMENTS

This research was supported in part by the National Science Foundation (TWC-1422018). Many students also helped contribute to this work, including Angelia, Arpita Agrawal, Chaiwut Chaianuchittrakul, Lu Chen, Jonathan Dinu, Jineet Doshi, Shawn Hanna, Brandon Jiang, Wen Shan Jiang, Stacy Kellner, Jennifer Kong, Raymond Li, Siddharth Nair, Alex Sciuto, Sarah Shy, Seo Yeon Sim, and Weijia Sun.